\documentclass[aps,prl,showpacs,twocolumn,superscriptaddress,showpacs,groupedaddress,nofootinbib]{revtex4-1}

\usepackage{graphicx}
\usepackage{dcolumn}
\usepackage{bm}

\usepackage{amssymb}   
\usepackage{amsmath,amssymb,amsthm,mathrsfs,amsfonts,dsfont} 

\usepackage{physics}   

\usepackage[colorlinks=true,citecolor=magenta, linkcolor=darkblue]{hyperref}
\usepackage[dvipsnames]{xcolor}
\definecolor{darkblue}{rgb}{0.0, 0.0, 0.55}

\pdfoutput=1

\begin{document}

\title{ Representations of  chiral alternative to vierbein  and
  construction of  non-Abelian instanton solution in curved
  space-time }  
\author{K. Maharana}
 \email{karmadevamaharana@gmail.com, karmadev@iopb.res.in}
\affiliation{Physics Department, Utkal University, 
 Bhubaneswar 751004, India}

\fontsize{10pt}{11.7pt}\selectfont

\begin{abstract}
A chiral alternative to the vierbein field 
in general relativity was considered by 't Hooft
in an attempt to facilitate the construction of a quantum theory
of gravity. These objects  $ f^a{}_{\mu\nu}$ behave like the
``cube root'' of the metric tensor. We try to construct specific 
representations of these tensors in terms of Dirac $\gamma $ 
matrices in Euclidean and Minkowski space, and promote these 
to curved space through Newman-Penrose
formalism. We conjecture that these  are new objects with physical 
significance and are the analog of Killing-Yano tensors. As an 
application, we try  to construct non-Abelian instanton like solutions in
curved space from the flat space   't Hooft tensors. The space part
of the tensors are decomposed to a product of Levi-Civita tensor and  flat 
space Dirac gamma matrices. The gamma matrices are promoted 
to the curved space with the help of vierbeins, as in  
separability of Dirac equation by Chandrasekhar, using Newman-Penrose formalism.
The curved space generalisation of the instanton solution $A_{\mu}^a $ is
now  constructed by substituting all  objects to their general covariant form.
\end{abstract}


\maketitle

A consistent and complete quantum formulation of gravity is yet to be
 realised\cite{Carlip}. 
Problems are encountered  with quantizing even pure gravity
\cite{Sagnotti1,Sagnotti2,van de Ven,Nicolai}.
The very existence of matter fields, and in particular the fermions, further
necessitates  their 
incorporation in a complete  quantisation programme. Unlike
tensors, there is no spinor
representation for the diffeomorphism invariance corresponding to the  GL(4)
general coordinate transformations of general relativity\cite{Weinberg}.
 The only way to 
incorporate spinors in general relativity is through the introduction 
of vierbeins, or tetrads, introduced by Weyl\cite{Weyl}.
A tetrad behaves like the ``square root'' of the metric tensor
$ g_{\mu \nu}$. 
The spinor, which is  `` square root'' of a vector or tensor of rank one,
can be incorporated into general relativity by using the tetrad.
The construction of both fermions and bosons out 
of spinors and their various combinations has been exploited through 
this approach.

The variation of the action with respect to tetrads, or their higher 
dimensional incarnations, and the spin connections 
gives the Einstein field equations. Such an action is the basis 
 for  quantum formulation of gravity in string theory, as well as in 
other covariant formulations, and also in the canonical approach.
The examples of application of tetrad formalism are the calculation of the metric
perturbation of black hole solutions, reflexion of electromagnetic
waves and Dirac particles in such geometries.\cite{Chandrasekhar}.
In the canonical approach to quantum gravity, tetrads appear ultimately 
through  Sen-Ashtekar
variables \cite{Ashtekar}.

Another class of objects which will interest us, closely connected to 
the vierbein, are the self-dual or antidual
solutions to the Yang-Mills equations of the gauge theories
in the form of monopoles, instantons, etc.\cite{BPST,'t HooftM,Polyakov}
 These objects appear naturally  in the  gauge formulation of any grand unified theory.
Although these objects are not yet seen in direct experiments,
the instanton solves the U(1)  problem of QCD, as well as, it may
contribute to the baryon violating process in the early
universe \cite{'t Hooft3,'t Hooft4} . 
 Topological objects,  which are analogues of monopoles,
have been created in lower dimensional systems in the form
of vortex  and other configurations.

It would be of interest to have a theory where the SU(3) of Quantum Chromo 
Dynamics and the SU(2) $\times$ U(1) of electroweak interactions
come out naturally in a manner similar to the Dirac's relativistic 
theory of electrons,  where the spin, magnetic 
moment and other
consequences, like the $\mathbf{L}\cdot \mathbf{S} $ coupling, emerge automatically.

 In an attempt to quantize gravity, 't Hooft has considered a 
 chiral alternative to vierbein \cite{'t Hooft0}. The
 Einstein-Hilbert action, constructed through these fileds, takes an elegant form. 
However,  imposition of the gauge condition and  the consequent 
constraint term  added to the Lagrangian  makes it
non-renormalizable.

 A  rare natural appearance of
an internal SU(3) symmetry occurs in this formulation. As it stands, this
SU(3) does not seem to be related to Quantum Chromodynamics.
However, it is tempting to speculate that the elementary fields
could be variously U(1),  SU(2), and SU(3) representations,
somehow,  arising naturally from the chiral alternative.
This chiral alternative to the vierbein was explored by 't Hooft
as an object at the deeper level with interesting characteristics
that reflect some properties of the dual solutions. 

Another  motivation for looking at
the chiral alternative to the vierbein is to analyse
the ``cube root of the metric tensor'', so as to obtain
some objects analogous to the Killing-Yano tensors.
Starting with a metric admitting  Killing tensors and 
its ``square-root'', i.e., Killing-Yano tensors, it leads to 
 discovery of hidden symmetries, 
exotic  supersymmetries, and conserved entities of the 
background. The  separability  of the 
Hamilton-Jacobi equations are also a consequence there of \cite{Floyd,Penrose,Carter1,Carter2,Page,Frolov2007,Krtous,Kubiznak2009, Frolov2017,Gibbons,Santillan}. 
So, it is   expected
that the chiral alternative being a ``cube root'' of the metric tensor would have similar
attributes leading  to enumeration of symmetries, 
conservation laws, and separability of the equation.

Though the formulated Lagrangian in terms of chiral vierbein is not renormalizable, the 
chiral alternative has many interesting features, such as,
emergence of new symmetry, self(anti)-duality, etc. In this paper, 
we first try to construct the representations of the chiral 
alternatives in terms of Dirac $\gamma$ matrices in curved space. 

{\it\bf{A chiral alternative:}}
Since chiral fermions and vie(r)(l)beins play a very important role
in both covariant and canonical formulations of quantum gravity,  as
well as in term of  loop, and string theory,
't Hooft has introduced an alternate new interesting object 
that behaves like the 
``cube root'' of the metric tensor, instead of the vierbein. Following
is a brief review  of the role of different fields leading to the
chiral alternative, and in setting up the basic formalism \cite{'t Hooft0}.

For introduction of Dirac field to general relativity, as well as an
alternative  to the metric tensor as the fundamental variable in the
covariant formalism, it is useful to
introduce the ``square root'' of the metric tensor $g_{\mu\nu} $; the
vierbein  field ${e^a}_\mu $,
\begin{align}
   {e^a}_\mu \;{e^a}_{\nu}  =  g_{\mu\nu}   . \label{eq:g}
\end{align}
Here $\mu $, $\nu $ are four vector indices and  $a$ represents 
``internal'' indices. 

Instead of ${e^a}_\mu $
't Hooft has introduced a field  $ {e^a}_{\mu\nu}$  in curved
space-time that takes the values ${\eta^a}_{\mu\nu}$ in a locally flat
coordinate frame. This ${\eta^a}_{\mu\nu}$ is the  the invariant
self-dual tensor of the monopole solution obtained by him\cite{'t HooftM},
\begin{align}
{{\eta}^a}_{\mu\nu} = - {{\eta}^a}_{\nu\mu} = {\epsilon}_{a\mu\nu} 
 + {\delta}_{a\mu}\; {\delta}_{\nu 4}  - {\delta}_{a\nu}\; {\delta}_{\mu 4}, 
\end{align}
where $ a = 1,2 $ or $3$, and $\epsilon $ is 3-dim Levi-Civita symbol.

 This new field $ {e^a}_{\mu\nu}$ satisfies 
\begin{align}
{\epsilon}_{abc} \; {{e}^a}_{\mu\nu} \; {e^b}_{\kappa\lambda}\;
{e^c}_{\rho\sigma} \; {\epsilon}^{\mu\nu\kappa\rho } = 24 \sqrt{g}\;
g_{\lambda\rho} ,\label{eq:sqrtg}
\end{align}
or better,
\begin{align}
{\epsilon}_{abc}\; {{f}^a}_{\mu\nu} {f^b}_{\kappa\lambda}
{f^c}_{\rho\sigma}  {\epsilon}^{\mu\nu\kappa\rho } = 24\;
g_{\lambda\rho} .
\end{align}
This is invariant under any transformation of the form
\begin{align}
{e^a}_{\mu\nu} \Rightarrow {S^a}_b \; {{e^b}_{\mu\nu}},
\end{align}
where ${S^a}_b \in $ SL(3) and 
 $\det\: S = 1$ for the Euclidean space . 
Analogous to vierbein field, one introduces an SL(3) connection field
${A^a}_{b\mu}$ by demanding
\begin{align}
D_\mu \ {e^a}_{\alpha\beta} =   \left( 
 {\partial}_\mu  {e^a}_{\alpha\beta} -
  {{\Gamma}^{\lambda}}_{\mu\alpha} \; {e^a}_{\lambda\beta}  
  - {{\Gamma}^{\lambda}}_{\mu\beta} \; {e^a}_{\alpha\lambda} \right. \nonumber \\
\left. + {A^{a}}_{b\mu}\; {e^b}_{\alpha\beta}  \right)
   = 0 . \label{eq:cA}
\end{align}
This leads to \cite{'t Hooft0}, 
\begin{align}
  {A^{a}}_{a\mu} = 0,    
\end{align}
 and a bilinear expression in $ {e^a}_{\alpha\beta}$ :  
\begin{align}
  K^{ab} = \frac{1}{8}\; {\epsilon}^{\alpha\beta\mu\nu} \;
  {e^a}_{\alpha\beta} \; {{e^b}_{\mu\nu}},
\end{align}
which has an inverse $K_{ab}$.

By redefining 
\begin{align}
 {e^a}_{\mu\nu} \; (\det K)^{-\frac{1}{9} } = {{f^a}_{\mu\nu} },
\end{align}
equation ({\ref{eq:sqrtg}}) can be rewritten as
\begin{align}
{\epsilon}_{abc}\; {{f}^a}_{\mu\nu} {f^b}_{\kappa\lambda}
{f^c}_{\rho\sigma}  {\epsilon}^{\mu\nu\kappa\rho } = 24\;
{g_{\lambda\rho}},
\end{align}
where $ {f^a}_{\mu\nu}$ is the chiral alternative. The   
Lagrangian now becomes
\begin{align}
  \mathcal{L}_{c.a.} = \frac { {\epsilon}^{\kappa\lambda\rho\sigma } } {32} \;
  {f^c}_{\kappa\lambda} \; {f^b}_{\rho\sigma}\; {F^a}_{c\mu\nu}\;
  {\epsilon}_{abd} \;{f^d}_{\alpha\beta} \;{{\epsilon}^{\mu\nu\alpha\beta}},
\end{align}
which is the analog of the Lagrangian with vierbein          
\begin{align}
\mathcal{L}_{vierbein} = \sqrt{g}\; R = \det(e) \; {F^{ab}}_{\mu\nu} \; {e^\mu}_a \;
 {e^\nu}_b .
\end{align}

For the case of chiral alternative 
\begin{align}
  {F^{a}}_{b\mu\nu} = {\partial}_\mu {A^{a}}_{b\nu} - {\partial}_\nu
 {A^{a}}_{b\mu} + {{\left[ A_\mu , A_\nu \right]}^{a}}_{b}  \nonumber \\
  =
 {\frac{1}{2}}\; {\epsilon}_{abd} \;{{\eta}^d}_{\lambda\alpha}\;
 R_{\lambda\alpha\mu\nu} .
\end{align}
This gives a relation between $f$ and the metric $g$, i.e.,
for ``cube root'' of $g_{\kappa\tau}$,
\begin{align}
 {\epsilon}_{abc}\; {\epsilon}^{\mu\nu\lambda\rho}\; f^a{}_{\mu\nu} \;
 f^b{}_{\lambda\kappa} \; f^c{}_{\rho\tau}  = 24\;\;
  {g_{\kappa\tau}}.  \label{eq:tH}
\end{align}

In Minkowski space, putting a reality condition
\begin{align}
 {{\hat{f}}^a}_{\mu\nu} =  {{\left({{{f}}^a}_{\mu\nu}\right)}^{\ast}}, 
\label{eq:su3}
\end{align}
converts the internal SL(3) to an SU(3) symmetry.

\vskip 0.1in
\noindent
{\it\bf{An attempt to obtain a representation in flat Euclidean space:}}
Here we try to construct  $ f^a{}_{\mu\nu}$ in terms 
of Dirac { $\gamma^\mu $}
matrices to get a relation related to ({\ref{eq:tH}}) for flat 
Euclidean space,
and also for  Minkowski space. Then, as for the generalisation of Dirac
equation to curved space, we can promote 
the  $\gamma^\mu $ matrices, and hence  $ f^a{}_{\mu\nu}$  to
the curved space.

The 't Hooft tensor, 
introduced in the  context of monopole-like solutions, is given by 
\begin{align}
  \eta^a{}_{\mu\nu} = \epsilon_{a\mu\nu} + g_{4\mu}\; g_{\nu a} -
  g_{4\nu}\; {g_{\mu a}},  
\end{align}
where $(a = 1,2,3)$ and it has the property of being antisymmetric 
and self-dual.

If we express $ f^a{}_{\mu\nu}$ as
\begin{align}
 f^a{}_{\mu\nu} = {\grave{ \sigma}}^a \; \gamma_5 \; {\sigma_{\mu\nu}}, \label{eq:fE}
 \end{align}
 with  $  {\grave{ \sigma}}^a        $  corresponding to an internal SU(2),
 and  the $\gamma^\mu$ matrices  in Pauli-Dirac representation satisfy \cite{Sakurai},
\begin{align}
\sigma_{\mu\nu} = \frac{1}{2i} \left[\gamma_\mu ,\gamma_\nu \right],
\nonumber \\
\gamma_\mu \gamma_\nu + \gamma_\nu \gamma_\mu =
2\delta_{\mu\nu},\nonumber \\
\gamma_5 = \gamma_1 \gamma_2 \gamma_3 \gamma_4 = \frac{1}{4}
\epsilon_{\mu\nu\lambda\rho} \;\gamma^\mu \gamma^\nu \gamma^\lambda
\gamma^\rho ,\nonumber \\
 {\gamma_\mu}^\dag = \gamma_\mu ,
\end{align}
then  $ f^a{}_{\mu\nu}$ is antisymmetric in $\mu$ and $\nu $ and is
anti-self-dual modulo $\gamma_5$. Here $ f^a{}_{\mu\nu}$ appears as an
object in the direct product of two spaces, the internal space indices
$a,b$, and $c$, which  may arise from some SU(2) and  the 4-space Greek indices
${\mu}, \nu \ldots $ etc. of Lorentz group or the SL(2,C), and hence do not have the deep significance of
$\eta^a{}_{\mu\nu}$ as in the  instanton or monopole solution.
For $\kappa = \tau $, one can show that
\begin{align}
\epsilon_{abc}\; \epsilon^{\mu\nu\lambda\rho}\; f^a{}_{\mu\nu}\;
f^b{}_{\lambda\kappa}\;f^c{}_{\rho\tau} \nonumber \\ 
  = \epsilon_{abc}\;
\epsilon^{\mu\nu\lambda\rho} \;\sigma^a \; \sigma^b \; \sigma^c \;\gamma_5
\gamma_\mu \gamma_\nu \; \gamma_5 \;\gamma_\lambda \gamma_\kappa\; \gamma_5\;
\gamma_\rho \gamma_\tau \nonumber \\
 = 6i\; I_2 \times 12\; I_4. 
\end{align}
This expression must vanish for $\kappa \neq \tau$. 

To make the right hand side vanish for $\kappa \neq  \tau $, we may
take either the trace or the determinant of $ \epsilon_{abc} \;
\epsilon^{\mu\nu\lambda\rho}\; f^a{}_{\mu\nu}\;
f^b{}_{\lambda\kappa}\;f^c{}_{\rho\tau} $, so that 

\begin{align}
 \tr \left(  \epsilon_{abc} \;
\epsilon^{\mu\nu\lambda\rho}\; f^a{}_{\mu\nu}\;
f^b{}_{\lambda\kappa}\;f^c{}_{\rho\tau} \right ) \propto {g_{\kappa\tau}}.
\end{align}
Here, the expression on r.h.s. means the value of that component. 
One can also use
\begin{align}
\epsilon^{\mu\nu\lambda\rho}\; f^a{}_{\mu\nu}\;
f^b{}_{\lambda\kappa}\;f^c{}_{\rho\tau}  \propto {g_{\kappa\tau}},
\end{align}
where $g_{\kappa\tau} $ is the Euclidean flat space metric.
We can also get the general relation
\begin{align}
 \epsilon_{abc}\; 
\epsilon^{\mu\nu\lambda\rho}\; f^a{}_{\mu\nu}
f^b{}_{\lambda\kappa}\;f^c{}_{\rho\tau} + ( \kappa \rightleftharpoons
\tau )  =         \nonumber  \\
 6i\; I_2 \times 12 \;I_4 \;{g_{\kappa\tau}}. \label{eq:tHE}
\end{align}

So the relation ({\ref{eq:tH}}) is not produced exactly, but
equation ({\ref{eq:tH}}) satisfies equation ({\ref{eq:tHE}}) modulo
a constant and identity matrices. 
Since we have considered in the simplest case $ \sigma^a $
to be two dimensional, and the space to be a direct product space,
it would be more appropriate to call our construct as an attempt 
to  a realisation, rather than
a representation. It is also intriguing that due to the presence of
$ \gamma$'s in equation 
({\ref{eq:fE}}), $  f^a{}_{\mu\nu}$   can act on spinors.

\vskip 0.1in
\noindent
{\it\bf{Minkowski space:}}
In the Minkowskian case,
we again define 
\begin{align}
f^a{}_{\mu\nu} = \sigma^a \; \gamma_5\; {\sigma_{\mu\nu}}, \label{eq:fgamma}
\end{align}
which is antisymmetric in $\mu$ and $\nu$, and is anti-self-dual modulo
$i\gamma_5 $. Here we use the $\gamma_\mu $ matrices of Bjorken and
Drell \cite{Bjorken},
\begin{align}   
\sigma_{\mu\nu} = \frac{1}{2i} \left[\gamma^\mu ,\gamma^\nu \right],
\nonumber \\ 
\gamma_\mu \gamma_\nu + \gamma_\nu \gamma_\mu =
2 g_{\mu\nu},\nonumber \\
 g_{\mu\nu} = ( 1, -1 , -1 , -1 ) \nonumber \\
\gamma_5 = -i \gamma_0 \gamma_1 \gamma_2  \gamma_3 .
\end{align}
 We get, for $\kappa =\tau$ ,
\begin{align}
\epsilon_{abc}\; \epsilon^{\mu\nu\lambda\rho}\; f^a{}_{\mu\nu}
f^b{}_{\lambda\kappa}\;f^c{}_{\rho\tau}  =  6i\; I_2 \times (-i)\;12
\;I_4\;
{g_{\kappa\tau}}. 
\end{align}
To make it vanish for $\kappa \neq \tau$, we consider as in the
Euclidean case, and obtain 
\begin{align}
\epsilon^{\mu\nu\lambda\rho} \;f^a{}_{\mu\nu}\;
f^b{}_{\lambda\kappa}\;f^c{}_{\rho\tau} + ( \kappa \rightleftharpoons
\tau )  = 144\; I_2 \times  I_4 \; {g_{\kappa\tau}},
\end{align}
or take the trace of the left hand side to obtain 
\begin{align}
 \tr \left(  \epsilon_{abc}\; 
\epsilon^{\mu\nu\lambda\rho}\; f^a{}_{\mu\nu}
f^b{}_{\lambda\kappa}\;f^c{}_{\rho\tau} \right ) \propto g_{\kappa\tau} .
\end{align}

 Another object can be constructed 
\begin{align}
\breve{f}^a{}_{\mu\nu} = \sigma^a \;\left( g_{\mu\mu} \; g_{\nu\nu}
+ i \gamma_5 \right) \;{\sigma_{\mu\nu}},
\end{align}
which is antisymmetric in $\mu$ and $\nu$ and is anti-self-dual. Its properties
are similar to that of $ f^a{}_{\mu\nu}$.
However, promoting this to curved space will have problem as both the
left and right side would contain the metric tensor $g_{\mu\nu}$,
unless we write 
\begin{align}
\breve{f}^a{}_{ij} = \sigma^a\; \left( 1
+ i \gamma_5 \right)\; \sigma_{ij}, \nonumber\\
\breve{f}^a{}_{i0} = \sigma^a \;\left(- 1
+ i \gamma_5 \right)\; \sigma_{i0}.
\end{align}

\vskip 0.1in
\noindent
{\it\bf{Gauge instanton in curved space:}}
Chandrasekhar \cite{Chandrasekhar} was able to show the separability of the Dirac
equation in curved space by making use of the Newman-Penrose 
formalism\cite{Newman}.
 For this, the constant 
Pauli matrices, { $\sigma_i$},
constituting the {$\gamma_{\mu}$} matrices are replaced by
their natural generalization
\begin{align}
\sigma^i{}_{A{B'}} = \frac{1}{\sqrt{2}} 
               \left| \begin{array}{cc}
                l^i         & m^i \\
                {\bar{m}}^i & n^i  \end{array} 
               \right|,     \label{eq:NP1}
\end{align}
where {\boldmath ${{l}, {m}, {\bar{m}},{n} }$} are the null 
basis vectors \cite{Chandrasekhar}. 
For the Kerr space-time the contravariant basis are
\begin{align}
l^i = \frac{1}{\Delta} (r^2 + a^2, +\Delta, 0, a) \qquad
n^i = \frac{1}{2\rho^2} (r^2 + a^2, -\Delta, 0, a)    \nonumber \\
m^i = \frac{1}{\sqrt{2} \bar{\rho}} (i a \sin \theta, 0, 1, i \csc
  \theta ), \qquad  \qquad \qquad
\end{align}
where $\Delta = r^2 -2 M r + a^2$, with $M$ the mass and $a$ the
angular momentum per unit mass of the black hole, and $\bar{\rho} = r a
\cos\theta$.
Now, the Dirac gamma matrices $\gamma^{\mu}$ , for  the curved space 
expression, become \cite{Futterman}
\begin{align}
\gamma^{\mu}  =  \sqrt{2} 
 \left| \begin{array}{cc}
                0        &  {\left( {\sigma}^{\mu A B'}  \right)}^T   \\
      {\left({\sigma}^{\mu }_{A B'} \right)}            &  0 \end{array} 
               \right|.    \label{eq:NPgamma}
\end{align}
This $\gamma^{\mu}$ is to be substituted in eqn.({\ref{eq:fgamma}}), for example ,
to obtain $ f^a{}_{\mu\nu}$. So the instanton solution in curved
space-time, on substituting all  objects to their general covariant
form,
becomes 
\begin{align}
A^a_\mu =   \frac{2}{g}   \frac{  {e}^a_{\mu\nu} {(x - x_0)}^\nu }
{ (x - x_0 )^2 + {\lambda}^2} ,
\end{align}
where $x_0$ is free due to translational invariance, and $\lambda$ is a
free scale parameter. Using equations ({\ref{eq:fgamma}}),
({\ref{eq:NP1}}), 
({\ref{eq:NPgamma}}),  and the appropriate substitution of the null
basis vectors for Kerr and other geometries, gives the gauge  instanton solutions
of  corresponding curved spaces.

\vskip 0.1in
\noindent
{\it\bf{Discussion:}}
Besides its appearance in the soliton solution, the 't Hooft tensor 
 $ f^a{}_{\mu\nu}$ itself has many interesting properties. 
The use of twistors in producing anti-self-dual solutions of the Yang-Mills
equations \cite{Ward} indicates that  twistors are related to 't Hooft tensor  $ f^a{}_{\mu\nu}$. 
It is remarkable  that an internal SU(3)  transformation of 
$ f^a{}_{\mu\nu}$ appears for the Minkowski space. 
It is interesting to note  that two other  rare instances of an SU(3),
that  arise
naturally, are the one as a consequence
of a dynamical symmetry implying  the existence of a Runge-Lenz type vector in
3-dimension harmonic
oscillator potential\cite{Schiff},  and the other 
from the  Dirac equation in fractal dimension\cite{Kerner}.
However, the  SU(3)  of chiral alternative  is not the consequence of a potential, and 
its origin is geometrical.
It is natural to expect conserved quantities arising out of the SU(3) 
symmetry of equation ({\ref{eq:su3}}).
It would be interesting to relate these to
the Killing-Yano type symmetries.
For matter coupled to gravity, where there is coupling to 
higher spin states, the acceleration
is expected to depend on higher powers of the four velocity
and this may give rise to new type of conserved quantities \cite{de Wit}.
Symmetries of spinning particles in Schwarzschild and Kerr-Newman type
space times have been analysed by Gibbons, Rietdijk and van Holten
\cite{Gibbons}. They found new nontrivial supersymmetry corresponding to
the Killing-Yano tensor for the black hole space-time. This also plays an 
important role in solving the Dirac equation in these black hole metrics.
The fermionic symmetries are generated by the square root of 
bosonic constants of motion other than the Hamiltonian \cite{de Wit,van
  Holten,Gibbons}. Therefore, one would like to conjecture that the ``cube
root'' $ f^a{}_{\mu\nu}$ would have analogous relation with the symmetries
and the  conserved quantities.

This Killing-'t Hooft tensor 
$ f^a{}_{\mu\nu}$
would be a new mathematically interesting object to analyse.
Establishing such a connection is expected to lead to a better
 understanding of the structure of space-time as well as the origin of
 the gauge symmetries.

\vskip 0.1in
\noindent
{\bf Acknowledgements:} 
 I am grateful to  G. 't Hooft for critical comments, and 
  thank  B. de  Wit and H. Nicolai for discussions.
Grants from Nederlandse Organisatie voor Wetenschappelijk Onderzoek
(NWO) and Max Planck Institute for  Gravitational Physics 
(Albert Einstein Institute) are acknowledged with gratitude.

\bibliographystyle{prsty}

\end{document}